\documentclass[aps,prl,twocolumn,showpacs,groupedaddress]{revtex4-1}
\usepackage{graphicx}
\usepackage{textcomp}
\usepackage{amsmath}
\usepackage{amssymb}

\begin{document}
\draft
\title{Photonic realization of the quantum Rabi model} 
\normalsize

\author{A.  Crespi $^{1,2}$, S. Longhi$^{1,2}$, and R. Osellame$^{2,1\ast}$}
\address{$^1$ Dipartimento di Fisica - Politecnico di Milano, Piazza Leonardo da Vinci 32, 20133 Milan, Italy}
\address{$^2$ Istituto di Fotonica e Nanotecnologie - Consiglio Nazionale delle Ricerche, Piazza Leonardo da Vinci 32, 20133 Milan, Italy}


%
\bigskip
\begin{abstract}
We realize a photonic analog simulator of the quantum Rabi model, based on light transport in femtosecond-laser-written waveguide superlattices, which provides an experimentally accessible testbed to explore the physics of light-matter interaction in the deep strong coupling regime.
Our optical setting  enables to visualize dynamical regimes not yet accessible in cavity or circuit quantum electrodynamics, such as
   bouncing of photon number wave packets in parity chains of Hilbert space.
\end{abstract}



\maketitle

The quantum Rabi model \cite{Haroche06,Knight}  is a milestone in the history of
quantum physics. 
It describes
the simplest interaction
between quantum light and matter, i.e. a
two-level atom coupled to a single quantized electromagnetic mode, and
applies to a great variety of physical systems, including cavity
and circuit quantum electrodynamics (QED), quantum
dots, polaronic physics and trapped ions.  In spite of its simplicity, the integrability of this model has been 
proven only recently \cite{Braak}, and the experimental exploration of some extreme dynamical regimes of the Rabi model is still missing. 
In most experimental conditions, such as in cavity QED, simultaneous creation or annihilation
of an excitation in both atom and cavity mode is extremely unlikely, and 
the coherent atom-field dynamics is described by the Jaynes-Cummings (JC) model \cite{JC}, which is obtained from the quantum Rabi model via the rotating-wave approximation (RWA). The JC model has been very successful to accurately predict
several  experimental phenomena, such as  the observation of the vacuum Rabi mode splitting in Alkali atoms \cite{EJC1} and
vacuum Rabi oscillations in Rydberg
atoms \cite{EJC2}. Recent achievements in circuit QED enabled to explore the 
ultrastrong coupling regime of light-matter interaction \cite{USC1,USC2,BlochSiegert}, in which effects of counter-rotating terms can not be neglected, but can still be taken into account by means of a perturbative approach.
This has led to the observation of the Bloch-Siegert shift, i.e. a noticeable shift in the resonance
frequency of the driven atom \cite{BlochSiegert}.  Recently, an even stronger coupling in the field-atom interaction, the so-called deep strong coupling (DSC) regime, has been theoretically investigated \cite{DSC}. This regime includes the highly counterintuitive simultaneous excitation
or de-excitation of both the atom and the field. Concepts like Rabi oscillations should be abandoned,  and new  extreme physical phenomena, such as bouncing of photon number wave packets along parity chains in Hilbert space, come into play \cite{DSC}. Although cutting-edge
experiments that can access this novel light-matter
coupling regime are desirable, the current status of cavity or circuit QED does not allow to reach the DSC regime yet. 
Creating tabletop analog systems to simulate  in the lab extreme dynamical regimes or phenomena in the matter, which are not accessible in an experiment, has attracted an increasing interest in recent years (see, for instance, \cite{Baluta,qw1,qw2} and references therein).  The possibility to simulate the DSC regime of light matter interaction, using either classical or quantum systems, has been recently proposed in Refs.\cite{Longhi2011,Solanoun}. In particular, in \cite{Longhi2011} it was shown that propagation of classical light beams in curved photonic superlattices provides a classical realization in Hilbert space of the quantum Rabi model which can access all kinds of light-matter
coupling regimes.\\
\begin{figure}[htb]
\centerline{\includegraphics[width=8.2cm]{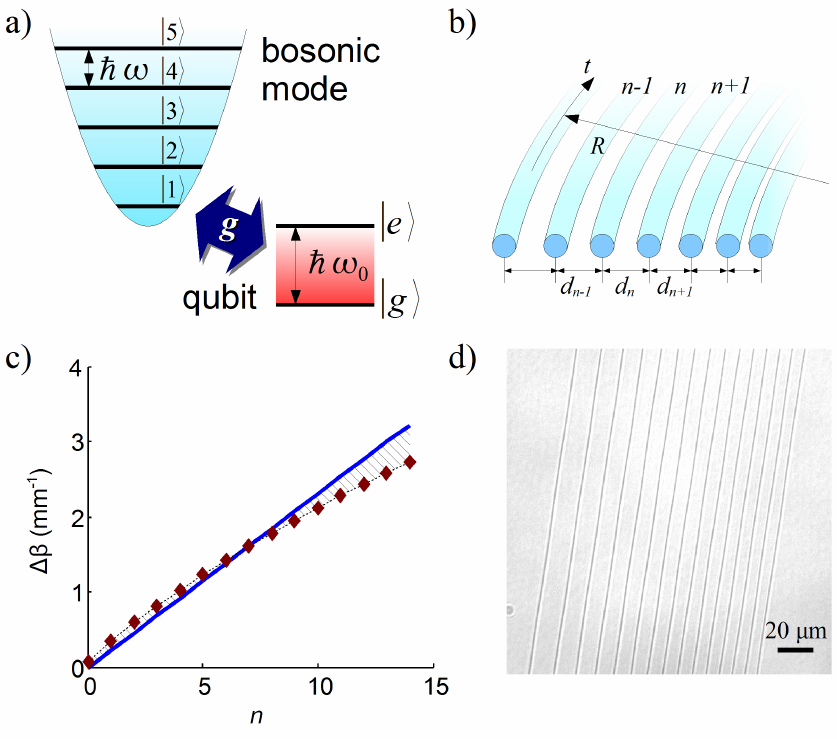}} \caption{
(Color online) (a) Schematic of a qubit interacting with a single
bosonic mode. (b) Photonic realization of the quantum Rabi model in Hilbert
space based on a semi-infinite curved binary waveguide array. (c) Variation $\Delta \beta$ in the propagation constants $\beta$ of the waveguides, induced by the axis curvature: actual trend without compensation due to a non-uniform waveguide spacing (diamonds) and target linear trend achieved after writing parameters compensation (solid line). (d) Top-view microscope image of the curved waveguide array.}
\label{fig1}
\end{figure}
In this Letter we realize a classical simulator of the quantum Rabi model, based on light transport in femtosecond-laser-written waveguide superlattices, and provide experimental evidence of the physics of light-matter interaction in the  DSC regime. \par
The Hamiltonian of the Rabi model, describing the interaction of a two-level atom (or a qubit) with a quantized field (or oscillator) (Fig.\ref{fig1}a), 
 reads \cite{Knight,Braak}
\begin{equation}
\hat{H}=(\hbar \omega_0 / 2) \hat{\sigma}_z+ \hbar \omega
\hat{a}^{\dag} \hat{a}+ \hbar g (\hat{\sigma}_+
+\hat{\sigma}_-)(\hat{a}+\hat{a}^{\dag}), \
\end{equation}
where $\hat{a}$ and $\hat{a}^{\dag}$ are the annihilation and
creation operators of the quantized oscillator with frequency
$\omega$, $\hat{\sigma}_z=|e \rangle \langle e|-|g \rangle \langle
g|$, $\hat{\sigma}_{+}=|e \rangle \langle g|$ and
$\hat{\sigma}_{-}=|g \rangle \langle e|$ are Pauli operators
associated to the two-level atom with ground state $|g\rangle$, excited state
$|e \rangle$, and transition frequency $\omega_0$, and $g$ is the
coupling strength. For a weak coupling $g \ll \omega_0$ and near-resonant excitation $\omega \simeq \omega_0$,  the
anti-resonant terms $\hat{\sigma}_+ \hat{a}^{\dag}$ and
$\hat{\sigma}_- \hat{a}$ entering in $\hat{H}$ can be neglected (RWA),  and the Rabi model reduces to the JC model of quantum optics \cite{Knight}. 
From a physical viewpoint, neglecting the anti-resonant terms means that processes corresponding to simultaneous excitation or de-excitation of the atom  
and of the quantized field are unlikely. In this regime,  the dynamics in Hilbert space, spanned
by the states $ |g \rangle |n \rangle$ and $|e \rangle |n \rangle$
describing $n$ quanta of the field with the atom in the ground or in
the excited state,  splits into an infinite
sequence of state doublets, and phenomena like vacuum Rabi oscillations or collapse and revival of populations can be observed. Without the RWA, the dynamics in the Hilbert space is more complex and  splits into two parity chains  \cite{DSC}.
In fact, let us expand the state vector $|\psi(t) \rangle$ of the atom-field system on the  $ \{ |g \rangle |n \rangle, |e \rangle |n \rangle \}$ basis as $|\psi(t) \rangle=\sum_{n=0}^{\infty} [a_n(t)
|e\rangle |n \rangle +b_n(t) |g \rangle |n \rangle]$, and let us
introduce the amplitudes $c_n(t)$ and $f_n(t)$, defined by
$c_n(t)=a_n \; , f_n(t)=b_n$  for $n$ even, and $c_n(t)=b_n \; ,
f_n(t)=a_n$ for $n$ odd. Then, the  temporal evolution of the
amplitudes $c_n$ and $f_n$ decouples into two parity chains as follows \cite{DSC}
\begin{eqnarray}
i \frac{dc_n}{dt} & = & \kappa_n c_{n+1}+\kappa_{n-1}c_{n-1}
+(-1)^n \frac{\omega_0}{2}c_n+n \omega c_n \;\;\;\;\; \\
i \frac{d f_n}{dt} & = & \kappa_{n} f_{n+1}+\kappa_{n-1} f_{n-1}
-(-1)^n \frac{\omega_0}{2}f_n+n \omega f_n  \;\;\;\;\;\;
\end{eqnarray}
($n=0,1,2,3,...$), where we have set
$\kappa_n=g \sqrt{n+1}$.  

\begin{figure}[htb]
\centerline{\includegraphics[width=8.2cm]{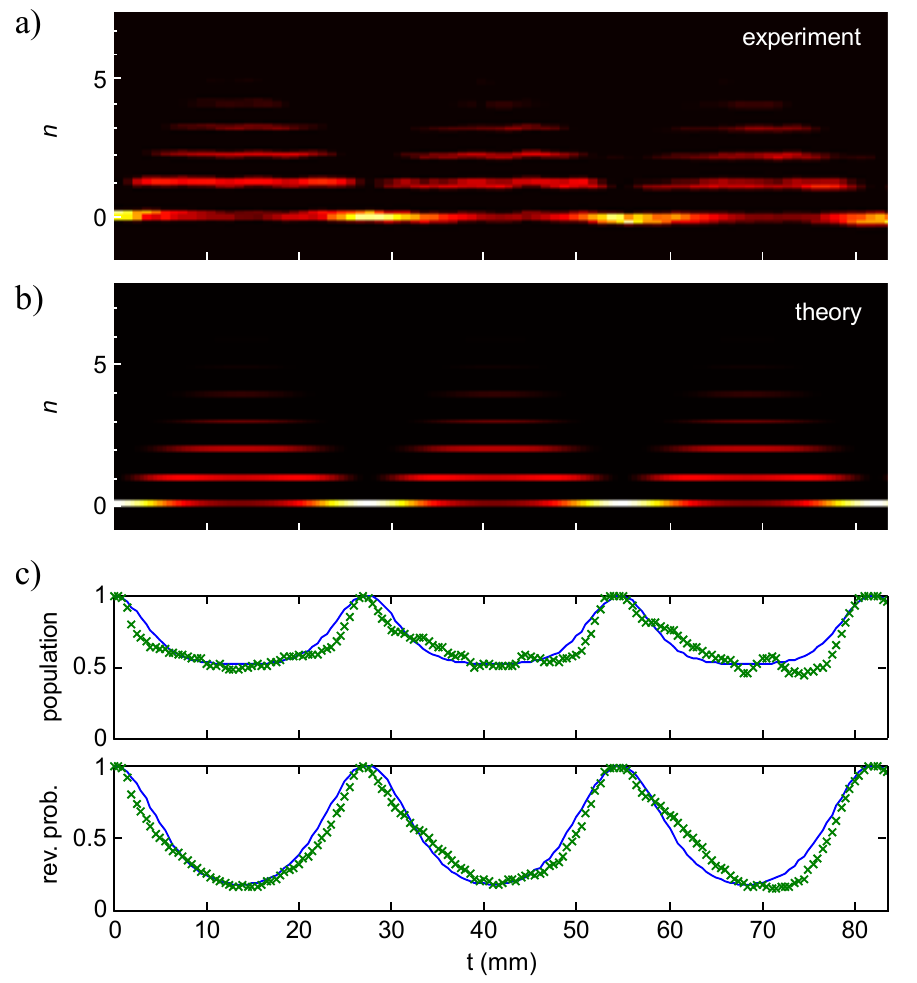}} \caption{
(Color online) (a) Measured and (b) predicted light intensity distributions in the waveguide array for $\omega_0=0$, showing the periodic bouncing of photon number statistics $P(n,t)$ in Hilbert space. (c) Evolution of population $P_e$ of level $|e \rangle$ (upper panel) and of the revival probability $P_r$  (lower panel) (crosses: experimental points; solid line: theoretical model).}
\label{fig2}
\end{figure}

\begin{figure*}[htb]
\centerline{\includegraphics[width=16.4cm]{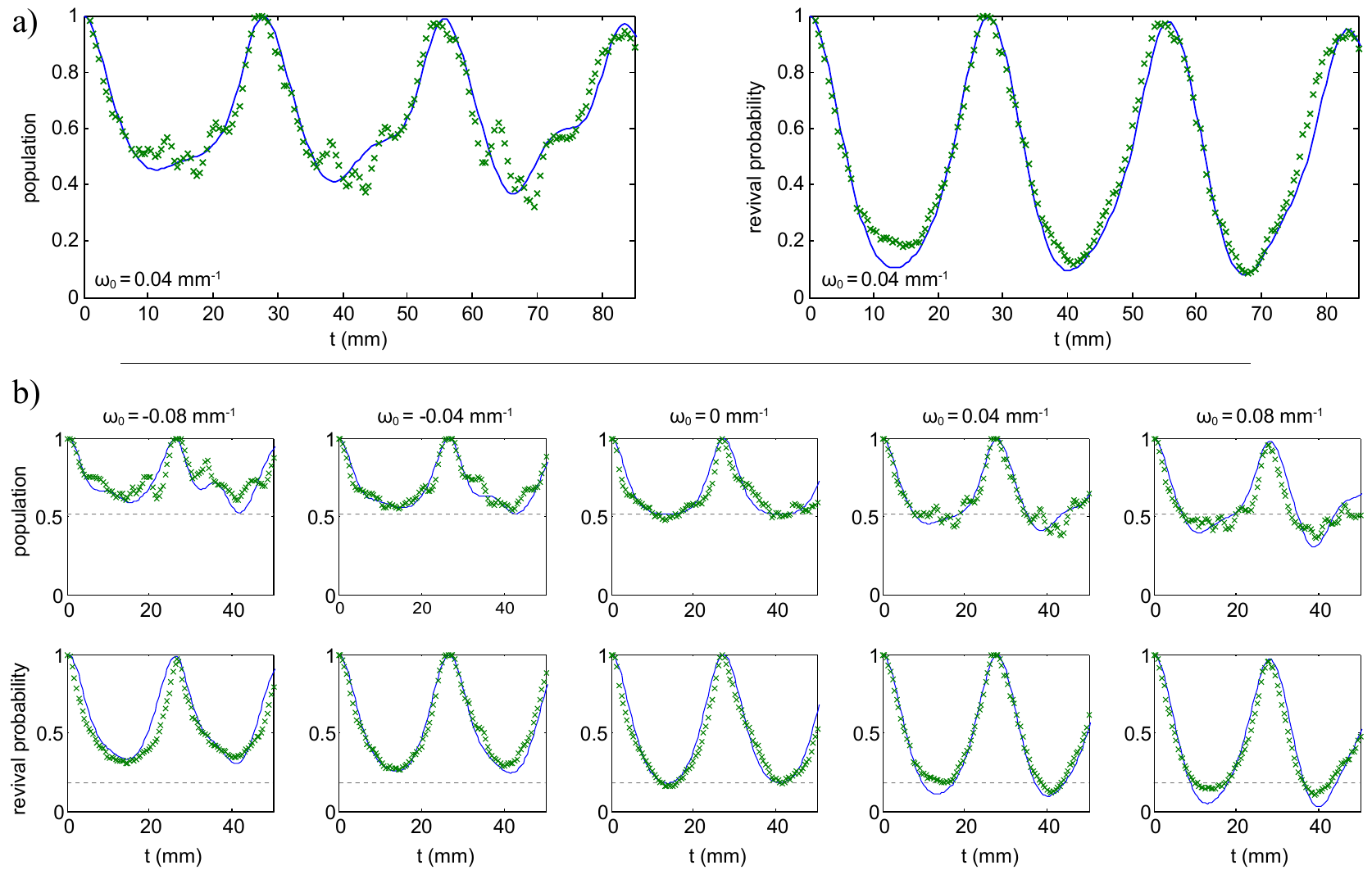}} \caption{
(Color online) (a) Same as Fig.2c, but for a waveguide array with $\omega_0$= 0.04 mm$^{-1}$. (b) Early evolution of revival probability $P_r$ and populations $P_e$ (for $\omega_0>0$) and $P_g=1-P_e$ (for $\omega_0<0$) as $\omega_0$ is tuned from negative (initial state $| \psi(0) \rangle=|g \rangle |0 \rangle$) to positive  (initial state $| \psi(0) \rangle=|e \rangle |0 \rangle$) values. Solid curves: theory; crosses: experimental points. The dashed  horizontal lines in the plots 
show, as reference levels, the minimum values of revival probability and populations attained in the first bounce interval for the $\omega_0=0$ case.}
\label{fig3}
\end{figure*}

As noticed in \cite{Longhi2011}, Eqs.(2) and (3) are analogous to the coupled-mode equations describing light transport in
two decoupled semi-infinite binary photonic lattices in the
tight-binding approximation with a superimposed transverse index
gradient and with non-uniform coupling constant $\kappa_n$ between
adjacent waveguides $n$ and $(n+1)$. In the optical analogue, the temporal evolution of the amplitude probabilities  $c_n$ and $f_n$ is mapped into spatial 
propagation of the light amplitudes in the various waveguides (see Fig.1b).  
The transition frequency $\omega_0$ of the two-level atom corresponds to the
propagation constant mismatch of alternating waveguides of the
superlattice, whereas the frequency $\omega$ of the quantized mode is defined by 
the waveguide propagation constant gradient superimposed to the superlattice. 
Note that the two parity chains of Eqs.(2) and (3) are simply obtained from each other by changing the sign of $\omega_0$.
The structure of the optical superlattice that realizes one of the two parity chains is shown in Fig.\ref{fig1}b. It comprises a semi-infinite planar array of curved waveguides with engineered distances $d_n$ in order to implement the desired coupling coefficients $\kappa_n$. The circular curvature of the waveguides introduces a transverse refractive index gradient leading to $\omega=2 \pi n_{\mathrm{eff},n} d_n/(R
\lambda)$, where $n_{\mathrm{eff},n}$ is the effective refractive index of the $n$-th waveguide,
$\lambda$ is the wavelength of light and $R$ the radius of curvature. Since the distances $d_n$ between different waveguides are not uniform in the array (Fig.\ref{fig1}b), the value of $\omega$ also varies with the waveguide number $n$ (see diamonds in Fig.\ref{fig1}c). To achieve a uniform $\omega$ in the array a compensation is required in the effective index $n_{\mathrm{eff},n}$ of the waveguides,  yielding a uniform $n_{\mathrm{eff},n} d_n$ product. In addition to this uniform transverse index gradient, the $\omega_0$ detuning is implemented by a further modulation of the effective refractive index in alternating waveguides. The present design of the photonic device differs from that proposed in \cite{Longhi2011}, where a three-dimensional layout was considered. In an actual implementation that geometry suffered from second-neighbours interactions and difficulties in simultaneous imaging of waveguides at different depths.

The photonic structure shown in Fig.\ref{fig1}b is manufactured by femtosecond laser waveguide writing on a fused silica substrate. The second harmonic of an Yb-based femtosecond laser (FemtoREGEN, HighQLaser GmbH), delivering 400~fs pulses, is used for the writing process. An optimal processing window was found at 20~kHz repetition rate, 300~nJ pulse energy and 10~mm/s translation speed. The laser beam is focused 170~$\mu$m below the glass surface by a 0.45~NA, 20$\times$ objective. Figure \ref{fig1}d shows a top-view microscope image of a typical manufactured array of 15 curved waveguides ($R$ = 650 mm). Note that the separation between the waveguides is not uniform, ranging from 6.6 $\mu$m to 14 $\mu$m. 
The modulation of the effective refractive index $n_{\mathrm{eff},n}$, requested for achieving both a uniform index gradient $\omega$ and the alternating mismatch $\omega_0$, is obtained by exploiting the refractive index change dependence of the guiding core on the writing speed. Calibration of this dependence provided a rate of $dn_{\mathrm{eff},n}/dv$ = 1.5$\times$10$^{-5}$ s mm$^{-1}$, thus requiring a variation of the writing speed in the range 10-14 mm s$^{-1}$ to achieve a uniform $\omega$ and a further modulation up to $\pm$0.3 mm s$^{-1}$ to implement the detuning $\omega_0$.

Femtosecond laser writing in fused silica creates color centers that provide fluorescent emission at about 650~nm when light at 633~nm is propagated in the waveguide \cite{Szameit2007apl}. Top-view imaging of the fluorescence signal is a convenient way to visualize and quantitatively estimate the light distribution along the waveguide array \cite{Szameit2010jpb}, rejecting the background light by a notch filter at 633~nm. 

In all the experiments presented below an array of 15 curved waveguides was manufactured with parameters $\omega$ = 0.23~mm$^{-1}$ and $g$ = 0.15~mm$^{-1}$. The condition $g \simeq \omega$ breaks the RWA and allows the investigation of the DSC regime. A first experiment addressed the limiting case $\omega_0 = 0$, which corresponds to degenerate qubit levels. In this case, $\hat{H}$ can be diagonalized by a Lang-Firsov transformation \cite{Lang}, and the energies turn out to be equally spaced by $\omega$, yielding a strictly periodic temporal dynamics of photon number distribution $P(n,t)$ and populations with period $T=2 \pi / \omega$. Figure \ref{fig2}a shows the top-view image of light distribution in the array, which provides the evolution of  $P(n,t)$. In the experiment, the $n$ = 0 waveguide is excited at the input plane (left side in Fig. 2a), corresponding to the system in the initial state $| \psi(0) \rangle = |e \rangle | 0 \rangle$. A periodic bouncing of the photon distribution can be clearly observed, in very good agreement with the expected one (Fig.\ref{fig2}b). The qubit population of the $|e \rangle$  level is given by $P_e(t)=\sum_{n=0}^{\infty} \left\vert c_{2n} \right\vert^2$, while the revival probability is defined as $P_r(t)=|\langle \psi(t)| \psi(0) \rangle|^2=\left\vert c_0 \right\vert^2$. The evolution of these two quantities can be readily retrieved from the light intensity map of Fig.\ref{fig2}a, and is depicted in Fig.\ref{fig2}c. Note that the measured behaviors of population and revival probability closely follow the theoretical curves,  with a period ${T}  \simeq  27.3$ mm. 

\begin{figure}[htb]
\centerline{\includegraphics[width=8.2cm]{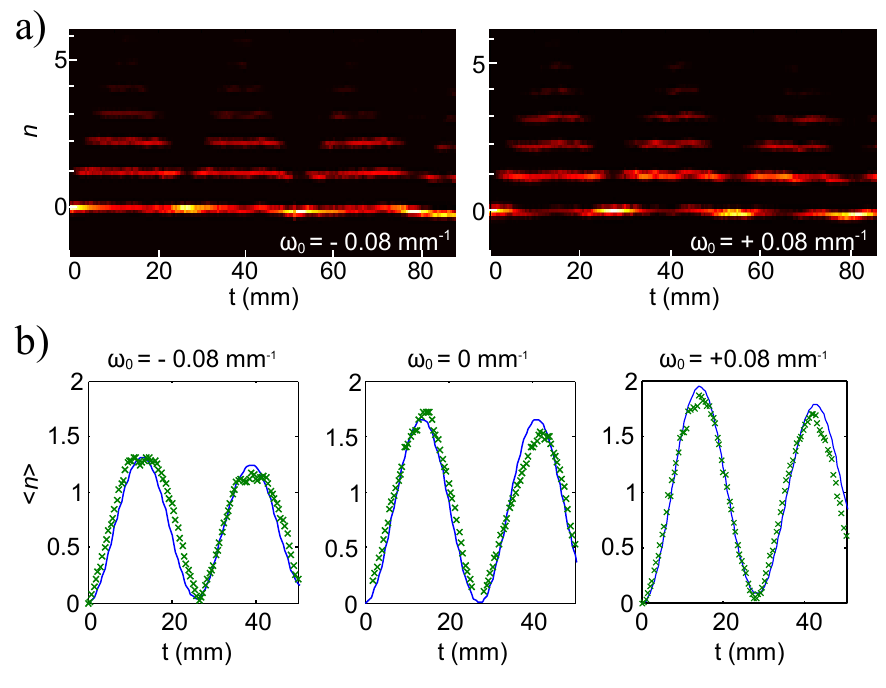}} \caption{
(Color online) (a) Measured light intensity distribution in a waveguide array for $\omega_0$ = -0.08 mm$^{-1}$ (left panel) and $\omega_0$ = +0.08 mm$^{-1}$ (right panel). (b) Evolution of the mean photon number $\langle n \rangle$ for $\omega_0$ = 0, $\pm$0.08 mm$^{-1}$.}
\label{fig4}
\end{figure}

When $\omega_0 \neq 0$, the energies of $\hat{H}$ are not equally spaced, and  the bouncing dynamics ceases to be periodic \cite{DSC}. However, for $\omega_0$ smaller than $\omega$, an imperfect bouncing dynamics can be still observed in the early stage of the evolution. As an example, Fig.\ref{fig3}a shows the evolution of $P_e$ and $P_r$ observed in an array with $\omega_0$ = 0.04 mm$^{-1}$, together with the corresponding theoretical curves, for the system in the initial state $|\psi(0) \rangle =|e \rangle |0 \rangle$. Smearing out of the periodic dynamics at successive bounces can be appreciated in this case. For $\omega_0 \neq 0$, the evolution of both revival probability and populations, besides losing the exact periodic behavior, is sensitive to the sign of $\omega_0$. In fact,  according to Eqs.(2) and (3) the sign of $\omega_0$ defines two distinct initial states of the qubit:  the qubit is in the excited state for $\omega_0>0$, whereas it is in the  ground state  for $\omega_0<0$.  Such different initial conditions excite two different  parity chains in Hilbert space, leading to different field-atom dynamics for the same initial photon number distribution. This is clearly shown in Fig.\ref{fig3}b, which depicts the early evolution of both revival probability $P_r$ and populations $P_e$ (for $\omega_0>0$) and $P_g=1-P_e$ (for $\omega_0<0$) as $\omega_0$ is tuned from negative to positive values, with no field quanta in the initial state. Note that, as $\omega_0$ is increased from negative to positive values, the minimum value of both the population and revival probability, reached within the first bounce, decreases. Such a behavior can be physically understood by observing that, when the qubit is initially in the $|g \rangle$ state (i.e. for $\omega_0<0$) and because of the absence of field quanta in the initial state,  the decoupled atom-field system is in its lowest energy state and excitation of field quanta and atom is thus less probable for an assigned coupling $g$ with respect to $\omega_0>0$. The possibility to create excitation in the atom and field starting from the initial state $|\psi(0)=|g \rangle |0 \rangle$, as demonstrated in the left panels of Fig. 3b corresponding to $\omega_0<0$, is a rather amazing prediction of the DSC regime of the Rabi model. Such a situation corresponds to creation of photons from an atom initially in its ground state which is strongly interacting with the vacuum state of the field.  In our ordinary wisdom of atom-field interaction, such a possibility would be obviously prevented because of energy conservation. In the DSC regime we actually do not violate energy conservation, because the energies of excitations ($\hbar \omega_0$ and $\hbar \omega$) are of the same order of magnitude or smaller than the "coupling" energy ($\hbar g$).  To better highlight the different dynamics of the coupled field-atom system  corresponding to the initial states $| \psi(0)=|g \rangle |0 \rangle$ and $|\psi(0)=|e \rangle |0 \rangle$, in 
Fig.\ref{fig4}  we show the detailed evolution of photon number distribution $P(n,t)$ for $\omega_0$ = -0.08 mm$^{-1}$ and $\omega_0$ = 0.08 mm$^{-1}$, together with the average number of created photons $ \langle n \rangle$. Note that in both cases $\langle n \rangle$ follows the bouncing dynamics of $P(n,t)$. The maximum value of $ \langle n \rangle$ within the bouncing cycle is 
noticeable larger in the $\omega_0>0$ case. This behavior is basically due to the fact that in this case the atom is initially in the excited state and such an extra energy is available to create photons.
 \par
In conclusion, we have experimentally investigated the physics of light-matter interaction in the deep strong coupling regime using a photonic analog simulator of the quantum Rabi model based on light transport in engineered waveguide superlattices realized by femtosecond laser writing. Our optical set up provides a simple tool  to visualize in the lab extreme dynamical regimes of light-matter interaction not yet accessible in cavity or circuit QED systems, and could be further extended to simulate non-Hermitian extensions of
the quantum Rabi model or related models \cite{Korsch} by introduction of optical gain and absorption in the guiding structure.

\par The authors acknowledge support by the italian MIUR (Grant No. PRIN-2008-YCAAK).\par
$^*$ Corresponding author: roberto.osellame@polimi.it

\end{document}